\documentclass[11pt]{article}
\usepackage[utf8]{inputenc}
\usepackage{natbib}
\usepackage{graphicx}
\usepackage[version=3]{mhchem} 
\usepackage{xcolor}
\usepackage[a4paper, portrait, margin=1in]{geometry}
\title{A Chemical Modelling Roadmap Linking\\ Protoplanetary Disks and Exoplanet Atmospheres\\\vspace{0.5cm}
  \small{Invited Review for the American Chemical Society's Special Issue:\\``Chemical Complexity in Planetary Systems''}}

\author{Christian Eistrup\\Email: eistrup@protonmail.com}

\begin{document}
\maketitle
\tableofcontents

\begin{abstract}

  Exoplanet atmospheres and protoplanetary disk chemistry are both active fields of research. Chronologically, they represent opposite ends of the process of planet formation: planets form in protoplanetary disks, and the composition of planets and their atmospheres should reflect the composition of the disk material they form from. The last decade of discoveries and insights by the ALMA observatory has provided a quantum leap in our understanding of the structures and chemical compositions of protoplanetary disks. Likewise, both ground and space-based facilities have provided peeks into the compositions of exoplanet atmospheres. However, the near future, with, first and foremost, the James Webb Space Telescope, and later also ESO's Extremely Large Telescope and ESA's Ariel mission are expected to provided constraints on exoplanet atmospheric composition with unprecedented precision, and also to shed new light on gas and ice compositions of protoplanetary disks. It is therefore timely to evaluate the current state of modelling efforts attempting to link protoplanetary disk chemistry and planet formation to exoplanet atmospheric compositions.

  The author has a background in astrochemistry related to planet formation. As such, this review discusses the differences between the simpler chemical approach of ``iceline chemistry'’ and the more chemically rigorous approach of utilising chemical kinetics. It is outlined which chemical effects may be at play in a planet-forming disk midplane, which effects are relevant under different conditions, and which tools are available for modelling chemical kinetics in a disk midplane. The review goes on to discuss some important efforts in the planet formation modelling community to treat chemical evolution, and, vice versa, efforts in the chemical modelling community to implement more physical effects related to planet formation into the chemical modelling.

  The aim of this review is both to outline some concepts related to planet formation chemistry, but also to encourage, not just collaboration between the planet formation modelling community and the astrochemical community, but also assistance and guidance from one community to the other. Guidance, regarding which effects, out of many, might be more relevant than others under certain planet formation conditions, and regarding why certain included effects lead to certain important modelling outcomes. As the research fields of exoplanet atmospheres and protoplanetary disks near new frontiers in observational insights with upcoming facilities, developing appropriate modelling frameworks (including physical and chemical effects) is paramount to ultimately enable the linking of a chemically characterised exoplanet atmospheres to its formation history in its natal protoplanetary disk.
\end{abstract}

\section{Introduction}
\label{intro}
Chemistry of planet formation regards the chemical journey from disk to planet for material in protoplanetary disks (prior to the formation of planets), how the evolution in these disks leads to the formation of planets within them, and how the material in the disks become incorporated into these planets. From this it follows that the chemical contents in planets and exoplanets observed today should match the contents of the material that went into forming the planets in their natal protoplanetary disks (PPDs).

The physical conditions in a PPD environment are not able to alter the relative amounts of the elements (elemental ratios), such as C/H, C/O, S/O, although the atomic and molecular carriers of elemental H, C, N, O, S and so forth may change during the process of planet formation. Thereby, the global (for an isolated young star and PPD) elemental ratios are constant. One example: an observed amount of water vapor in the atmosphere of an exoplanet does not imply that the material forming the planet had the same amount of water, since water molecules can be produced and destroyed during the planet formation process, and so there will be different carriers of elemental oxygen during the process. However, the total amount of elemental oxygen in an exoplanet atmosphere must match the total amount of oxygen contained in the planet-forming material, since oxygen atoms cannot be produced or destroyed under the condition found during the process of planet formation. From this follows that an observed ratio of, say, carbon and oxygen contained in an exoplanet must match the ratio of carbon to oxygen in the material that formed the planet, which may be a combination of gas, refractory silicate and/or carbonaceous rock, and volatile ices.

One important caveat in this picture is that homonuclear species, such as \ce{H2} gas, \ce{N2} gas and \ce{O2} gas are unobservable under protoplanetary disk conditions, since these molecules do not have permanent dipole moments. For this reason, the elemental H, N and O that these species may carry, is not accounted for in the observational evidence, and this introduces some errors in the measured elemental ratios. This may, to some degree, be circumvented by observing isotopologues of these species, like HD instead of \ce{H2} \citep[see, e.g.,][]{trapman2017hd}, \ce{^14N^15N} instead of \ce{^14N^14N}, and \ce{^16O^18O} instead of \ce{^16O^16O} \citep[see, e.g.][]{goldsmith2011}. However, this approach requires insights into, and assumptions about the ratios of D/H, \ce{^14N}/\ce{^15N}, \ce{^18O}/\ce{^16O}, in order to recover the \ce{H2}, \ce{N2} and \ce{O2} abundances. This, in turn, means that the recovered abundances will carry the uncertainties associated with these assumptions.

\begin{figure}[h!]
    \centering
    \includegraphics[scale=0.22]{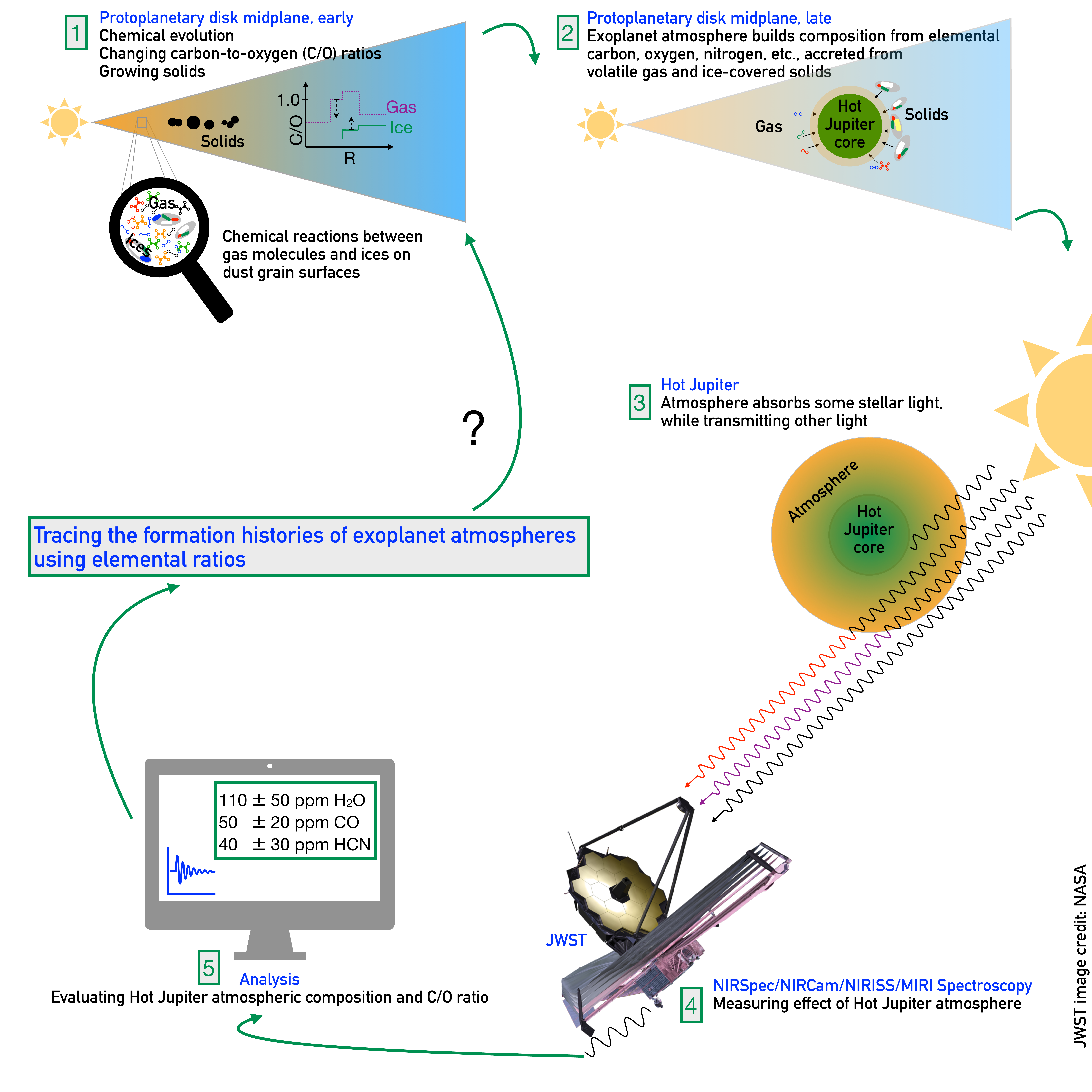}
    \caption{A chronological overview of the effects of planet formation chemistry, from the onset of planet formation in a protoplanetary disk (1) to analysis of JWST data from observing programs that characterise exoplanet atmospheric compositions (4-5). In this illustration with a focus on Hot Jupiter exoplanets. This illustration was made by the author. JWST illustration credit: NASA.}
    \label{nasa}
\end{figure}

Disk chemistry leads into planet chemistry, as planets are made in disks, between 100kyr and 10 Myr (see review by \citet{williams11}) after the formation of the disk. As such, in order to understand the chemistry of planet formation, understanding the process of planet formation is needed. The question is how material found in a disk forms planet, and then, in turn, how the chemical composition of this material changes in time. Additionally, assumptions need to be made regarding which components of the disk midplane are assumed to contribute to forming an exoplanet atmospheres that can be chemically characterised. In this context, several types of atmospheric accretion should be considered. One is that a forming planet can accrete midplane gas onto its surface and have this contribute material to its atmosphere, which adds the local gas composition to the exoplanet atmospheric composition. Another is that solid material (dust grains, pebbles and large bodies) may be accreted \citep[see, e.g.][]{mordasini2016}, as they drift dynamically from outer to the inner parts of the disk midplane \citep[see, e.g.][]{bitsch2015,bitsch2019pebblesgiants,trapman2019}, carrying with them chemical species in ice form. These ices may sublimate and add material to the composition of exoplanet atmospheres after the solids are accreted by the forming planet. And the composition of these ices may, in turn, be different from the composition of the local disk midplane gas accreted onto the planet. These chemical effects should be considered when attempting to model the chemical compositions of forming planets. An illustration of how chemistry plays a role throughout the formation of a Hot Jupiter atmosphere is shown in Fig. \ref{nasa}.

Additionally, observational evidence indicates that material from the upper layers of a PPD may ``fall'' directly down to the midplane through gaps in disk induced by effects of planet formation, in so-called meridional flows, see \citet{teague2019waterfall}. This phenomenon was investigated through physical and chemical modelling by \citet{cridland2020vert}, and shown to have effects on resulting chemical composition of the forming planet. In other words: planets accrete material of various sorts from various places to form their atmospheres, and it is therefore imperative to understand the chemical evolution and compositions of all these types of PPD materials, and be able to model a planetary atmosphere, assuming it to be formed from a mix of these materials..

After ten years of discoveries and insights into the structures \citep[see overview of DSHARP program and review by][]{andrews2018dsharpoverview,andrews2020review} and chemical compositions \citep[see, e.g., overview of MAPS program by][]{oberg2021maps} of PPDs done by the Atacama Large (sub)Milliter Array in Chile, as well as increasingly detailed knowledge about the chemical compositions of exoplanet atmospheres provided by current state-of-the-art observational facilities (and with the upcoming facilities such as the James Webb Space Telescope, ESA's Ariel mission and ESO's Extremely Large Telescope expected to break new ground), chemical connections between disks and planets are beginning to be drawn. This review will outline some of the research that has been carried out to link chemistry between disks and planets. It will explain some of the concepts that are used, and address some topics of current discussion.

This review was written independently of the recent book chapter titled ``Chemistry During the Gas-rich Stage of Planet
Formation'' by \citet{bergincleeves2018book} from \emph{Handbook of Exoplanets}\citep{deeg2018exohandbook}, and the review titled ``Astrochemistry associated with planet formation'' by \citet{dishoeck2020planetform}. While there is some overlap between them, this current review has a heavy focus on chemical modelling of volatile chemistry in relation to planet formation in the disk midplane, and this makes it largely complementary to the previous review and book chapter. The aim here is to provide researchers in the fields of planet formation (modelling and observations) and exoplanet atmospheric characterisation (modelling and observations) with an overview of what chemical reactions in planet-forming PPDs mean, how these reaction may affect chemical evolution, and how such chemical evolution may be utilised in, and implemented into planet formation models.

\section{A classical picture of protoplanetary disk midplane chemistry}
\label{classic}
\begin{figure}
    \centering
    \includegraphics[scale=1]{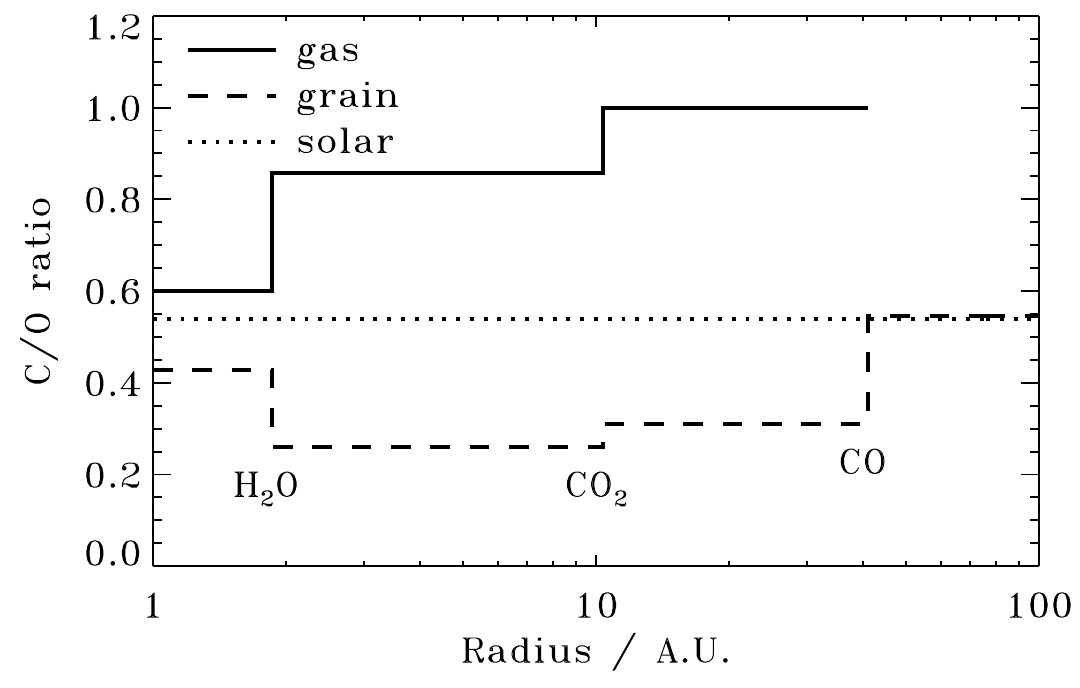}
    \caption{Carbon-to-oxygen ratios in gas and on grain surfaces (ice) in the pre-solar nebula as a function a disk midplane radius, with the solar ratio of 0.55 as a reference. See text for description. Reproduced from Fig. 1, \citet{oberg2011co}. Copyright 2011 Astrophysical Journal.}
    \label{obergco}
\end{figure}

One baseline for describing planet formation chemistry that has gained popularity in the fields of both exoplanet and disk research is the carbon-to-oxygen ratio diagram. This was first conceptualised in \citet{oberg2011co}, in which Fig. \ref{obergco} was published.

This diagram shows the carbon-to-oxygen ratios in gas and on grain surfaces (ice) in the pre-solar nebula, with the solar ratio of 0.55 as a reference. The ``steps'' in both the gas and the grain profiles has lead to this figure being referred to as a C/O step function, and these steps coincide with the disk midplane radii and associated temperature and density conditions that correlate with each of the shown volatile species transitioning from the gas to the ice-phase. Such a radius of transition is usually referred to as either a snowline, iceline or frostline. These terms are used somewhat interchangeably in the literature, although snowline and frostline are generally adopted when referring to the transition radius of \ce{H2O}, specifically. The term iceline is used, e.g., in \citet{eistrup2016,eistrup2018} as a reference to gas-to-ice transition radii for volatile molecular species generally, by referring to, say, the \ce{H2O} iceline or the CO iceline.

As seen in Fig. \ref{obergco}, these transitions cause abrupt radial changes to the profiles. This picture assumes that all of each molecular species changes from one radius/temperature to the next. It also assumes that not chemical reaction take place between any chemical species. There are five reservoirs of carbon and/or oxygen in this framework: \ce{H2O}, \ce{CO}, \ce{CO2}, carbonate grains and silicate grains \citep[see Table 1 in][for more detail]{oberg2011co}. This means that some carbon and oxygen is locked in refractory forms which do not transition between gas and ice in the diagram.

\subsection{Planet formation chemistry using icelines}

One novelty and appeal of Fig. \ref{obergco} was that it shifted away from a focus on individual molecular species to a focus on the ratio of carbon and oxygen as elementary constituents of these species, and the figure outlines where in the disk the gas and the solids have a higher or lower C/O ratio than the global (solar) value. This was an important shift. The physical conditions present in a PPD cannot alter the overall ratio of carbon to oxygen, but the ratio of individual molecules like, say, \ce{H2O}-to-\ce{CO2}, can be altered through chemical reactions. That means that tracing the incorporation of \ce{H2O} and \ce{CO2} into a planet provides less insight than tracing the incorporation of the elemental constituents (oxygen, hydrogen and carbon) into the planet. Fig. \ref{obergco} provides an easy-to-use framework for such an exercise. If an assumption is made about where in a disk midplane a planet's (primary) atmosphere is (mainly) accreted from, and an assumption about which type of material (gas or solids) is (mainly) accreted to form such an atmosphere, then is is straightforward to derive the C/O ratio (and similar ratios for other elements) of the resulting planetary atmosphere. This is, to zeroth order, simply a matter of reading off the C/O ratio value in the Fig. 1 diagram.

This has obvious links to the C/O ratios that have been observationally constrained using methods like atmospheric retrieval of exoplanets \citep{madhu2012,molliere2019,molliere2020}, as these atmospheric C/O ratios can be used to infer where in their natal protoplanetary disks these measured exoplanets once formed, thus providing a chemical connection to their formation histories. Another appeal of Fig. \ref{obergco}, which has high relevance for cometary science, is that it outlines where specific molecular species are in the gas and ice phases. Because \ce{H2O} has a high molecular binding energy ($E_{\mathrm{bin}}$=5770 K) corresponding to an iceline location coinciding with a midplane temperature of 120-150 K (dependent on the density of a given disk midplane), \ce{H2O} is in the ice phase in a larger part of the disk midplane than is, say, CO, which has $E_{\mathrm{bin}}$=855 K, and an iceline at midplane temperature of $\sim 20$K. For this reason, comets have a larger region of a disk where they can accrete icy solid material containing \ce{H2O} ice from than where they can accrete CO ice from. This may, in turn, have implications for cometary compositions \citep[see e.g.][and references therein]{eistrup2019comet}.

This review focuses more on carbon and oxygen-carrying species, rather than species carrying, e.g., nitrogen and sulphur. This is because the focus in the planet formation and PPD astrochemical community has largely been on the C/O ratio in disks and in exoplanet atmospheres, rather than similar ratios for N and S. The reason behind this is likely (at least in part) that C and O are more abundant than N and S, and that some commonly observed and modelled volatile species are carriers of C or O or both (e.g. \ce{CO}, \ce{CO2}, \ce{CH3OH}, \ce{H2CO}, \ce{H2O} and \ce{CH4}). However, \citet{turrini2021}, for example, modelled planet formation chemistry with a focus of N and S instead of C and O, and discussed how N and S could potentially be better tracers of planet formation history than C and O, when connecting characterised exoplanet atmospheres to PPD midplane conditions.

\subsection{From iceline \emph{radius} to iceline \emph{region} and ice/snow \emph{surface}}
\label{icelinedef}
The term ``iceline'' was explained in Section \ref{classic}. However, icelines are defined differently across the scientific literature on PPDs. Below is outline of some of these definitions, along with their benefits and drawbacks.
\begin{enumerate}
    \item Definition of iceline \emph{radius} by midplane \textbf{gas temperature} alone. This approach defines the iceline as a sharp transition from ice to gas moving inwards in the disk midplane (same step function framework as was shown in Fig. \ref{obergco}), with a specific gas temperature threshold in the midplane setting this transition. In this framework, all of a given molecule is assumed to change phases from ice to gas by ice molecules desorbing off the surfaces of dust grains and into the gas, from one radius to the next going inwards (with an associated increase in temperature). Vice versa, is the change from gas to ice phase (freeze-out) for a molecule when moving outwards in the disk, with decreasing temperatures. In this framework, all chemical species have iceline temperatures assigned to them, and given a temperature structure of a disk midplane, icelines for the different species can be set at the radius with the temperature closest to the species's iceline temperature.

    In addition to the original work by \citet{oberg2011co}, many other works have utilised this approach, amongst them \citet{madhu2012,alidib2014,bitschsavvidou2021}. It is noted that the choice of, and motivation for iceline temperature varies across studies. In \citep{bitschsavvidou2021} the condensation temperatures for varies molecules were adopted from \citet{lodders2003}, which meant that their iceline radii for the molecules depended only on one temperature per molecule, ignoring other physical effects. \citet{oberg2011co} adopted a range of temperatures for each molecule, stemming from the dependence of iceline radius on local gas and dust volume density, as prescribed by \citet{hollenbach2009}. Furthermore, \citet{oberg2011co} highlighted that a given molecule in the ice may feature varying desorption temperatures, dependent on such effects as trapping in the ice matrix (CO being trapped in an \ce{H2O} ice matrix, which, however, was shown by \citet{fayolle2011} to only trap 5-10\% of CO). Other effects that may cause binding energies of molecules to vary is whether or not, for example, a CO molecule is bound to a rocky (silicate grain) surface directly, or if the CO molecule is bound to other ice molecules that already reside on top of the bare rocky surface. Furthermore, in the case that the CO molecule is bound to other ices, whether these others ices are made up of CO as well, or possibly other molecules, see \citet{cleeves2014water}. Collectively, these effect therefore introduce uncertainties into the choice of a iceline temperature, especially if only one temperature is considered for each molecule.

    This approach, by construction, benefits from a high radial resolution of the midplane temperature structure, so as to have radii with temperatures as close as possible to the iceline temperatures of the considered molecules. Along the same lines, this approach is easy to implement, given knowledge about the gas temperature structure in the disk midplane. However, this approach is simplistic in its sole dependence on midplane gas temperature, as will be outline under Point 3 in this list.

    \item Definition of iceline \emph{region} by \textbf{gas temperature} \emph{and} \textbf{amount of ice to desorb}. This approach uses a single iceline/freeze-out gas temperature in the disk midplane the same way as in Point 1, but with the added physical consideration (for solid bodies drifting radially inwards in the disk), that more layers (monolayers) of ice on the surfaces of grains require more time to desorp. This means that even as a grain with ice on its surface crosses, say, the \ce{H2O} iceline as the grain drifts inwards, it does not immediately lose all \ce{H2O} to the gas. This leads to a radial region inside the temperature-defined \ce{H2O} iceline, in which some \ce{H2O} has desorbed, and some is still in the ice waiting to desorb. This regions therefore has \ce{H2O} in both the gas and the ice phases. The size of this region depends on factors such as the sizes of the drifting dust grains, their drifting speeds, the composition and the rate of change of the gas temperature moving inwards (the midplane temperature structure).

    Such iceline \emph{regions} were modelled and discussed by, e.g., \citet{piso2016}. See Fig. \ref{pisoicelines} for impression of the iceline regions from this paper, and Fig. \ref{obergdrift} for a cartoon depiction of how CO-ice-covered grains grow to form pebbles that drift across the CO iceline. The timescales for drift versus desorption of ices were discussed by \citet{piso2015}. Figure 3 in that study shows that for the fastest-drifting particles, the drift timescale is 1-2 orders of magnitude faster than the desorption timescale, with a desorption timescale of $\sim$1000 years for \ce{H2O} ice. However, for particles 10 times biggest or smaller than the fastest drifting ones, the timescales for drift and ice desorption are comparable (at 10k -- 1M years, depending on the ice molecule in question).
    \begin{figure}
    \centering
    \includegraphics[scale=0.5]{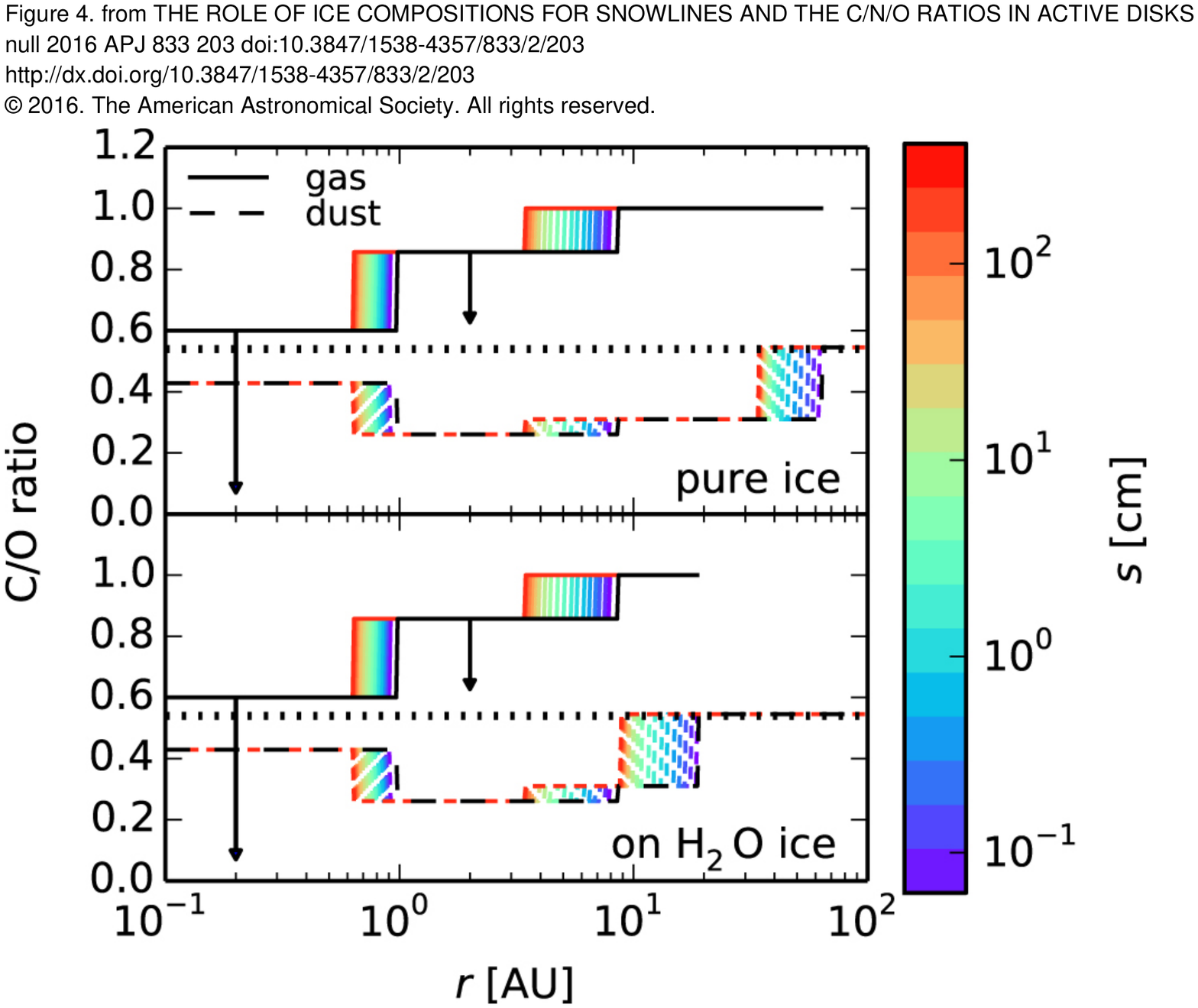}
    \caption{C/O ratios in gas and ice as a function of disk midplane radius. Iceline radii becomes iceline regions, when dependencies on ice compositions and sizes of drifting grains are accounted for. Reproduced from Fig. 4, \citet{piso2016}. Copyright 2016 Astrophysical Journal.}
    \label{pisoicelines}
\end{figure}
    \item Definition of iceline radius \emph{and} region by \textbf{balancing} the chemical effects of \textbf{freeze-out} of molecules from the gas onto grain surfaces, \emph{and} \textbf{desorption} of ice molecules off of grain surfaces into the gas (assuming a constant number of sites on grain surfaces, per volume, where ice molecules can reside, which can be achieved by assuming spherical grains of identical, and constant size, and a constant grain number density). Here, the molecules are said to be in the ice, if they are residing on the surfaces of the grains. In this framework, the midplane temperature, gas density, and grain number density are taken into consideration. Other factors involved in the framework include the mass of each molecular species, and the amount of monolayers of ice residing on the grain surface (please refer to \citet{cuppen2017review} for further details about monolayers on grain surfaces). Below is a brief discussion freely adopted from the review by \citet{cuppen2017review} (Eqs. 2-6) about how to compute the rate of freezing out for a molecule (or the ``rate of neutral grain accretion'', as it is also referred to as), and the rate of desorbing it off of a grain surface.

    The accretion rate $f_{\mathrm{acc,A}}$ for chemical species A from the gas onto a grain surface is:
    \begin{equation}
         f_{\mathrm{{acc},A}}=S_\mathrm{A}n_\mathrm{grain}\pi r^2_\mathrm{grain}n_\mathrm{gas}\mathrm{(A)}\times \sqrt{\frac{8k_\mathrm{B}T_\mathrm{gas}}{\pi m_\mathrm{A}}} \left[\mathrm{cm}^{-3}\mathrm{s}^{-1}\right],
         \label{acc}
    \end{equation}
    where $S_\mathrm{A}$ is the sticking efficiency of species A to the grain surface, $n_\mathrm{grain}$ is the number density of dust grains, $r_\mathrm{grain}$ is the radius of the grains (which are assumed to be spherical and one-sized), $n_\mathrm{gas}\mathrm{(A)}$ is the number density of species A in the gas phase, $k_\mathrm{B}$ is Boltzmann's constant, $T_\mathrm{gas}$ is the gas temperature, and $m_\mathrm{A}$ is the molecular mass of species A.

    The desorption $f_{\mathrm{des,A}}$, in turn, for chemical species A from the ice phase on a grain surface into the gas phase is:

    \begin{equation}
        f_{\mathrm{des,A}}=\sqrt{\frac{2N_sE_{\mathrm{bind,A}}}{\pi^2m_A}}\times \mathrm{exp}\left(-\frac{E_{\mathrm{bind,A}}}{kT_{\mathrm{grain}}}\right)\times n_\mathrm{s,A}\left[\mathrm{cm}^{-3}\mathrm{s}^{-1}\right],
        \label{des}
    \end{equation}
    where $N_s$ is the surface density of binding sites on the grain surface from which species A can desorb, $E_{\mathrm{bind,A}}$ is the kinetic binding energy of species A, $T_{\mathrm{grain}}$ is temperature of the dust grain, and $n_\mathrm{s,A}$ is the number density of species A on the grain surface, per unit volume. When modelling PPD midplanes, it is common to assume that the gas and the dust is well-coupled thermally (meaning that collisions between gas molecules and grains occur frequently enough as to ensure they stay the same temperature), which means setting $T_{\mathrm{grain}}=T_\mathrm{gas}$.

    Chemical kinetics codes that account for the chemical effects of freeze-out and desorption (as well as chemical reactions) determine the iceline radius or region of a given species based on balancing Eqs. \ref{acc} and \ref{des}. Based on the inputs to these formulae, the codes will compute a midplane radius with corresponding physical conditions, inside of which $f_{\mathrm{des,A}}$ is the larger term, and outside of which $f_{\mathrm{acc,A}}$ is larger. This radius will then usually be denoted as the iceline radius (it is noted that while Eqs. \ref{acc} and \ref{des} calculate the rates of accretion and desorption, respectively, of species A, and not the absolute abundances of the species in the gas and ice phases, the timescale of abundance equilibration at the radius where the two terms are equal is very short: \textless1 year at the radius of the \ce{H2O} iceline and \textless10 years at the radius of the CO iceline).

    However, both accretion and desorption are still active and at play in all parts of the disk midplane. In the inner disk, the accretion term is usually negligibly, and likewise for the desorption term in the outer disk. However, both effects being at play does mean that despite an iceline radius being assigned to species A, there is still a radial region just inside and outside this radius where neither of the terms calculated for accretion and desorption are negligible, even if one is the larger. This leads, in the chemical kinetics picture, to a small (dependent on the temperature structure) radial region where, at a given radius within, some amount of species A are in the gas phase, and some is ice on the grain surface, simultaneously. From Fig. 2.a in \citet{eistrup2016}, and assuming the disk midplane temperature structure from that study, the size of this radial region is $\sim$0.2 AU for \ce{H2O} in the inner disk, and $\sim$2-5 AU for CO in the outer disk. In other words, in the chemical kinetics picture, the phase change from gas to ice for species A is gradual with increasing midplane radius, rather than abrupt, and at a given radius some ratio of the the total amount of species A in the gas phase, and the rest is in the ice. This was depicted in \citet{eistrup2016}, Fig. 2.a., and it results in a C/O ratio diagram, which differs slightly from Fig. \ref{obergco}, and can be seen in Fig. \ref{eistrupco}.

    The chemical kinetics method of balancing the accretion and desorption rate terms means that the iceline radius depends on more than midplane gas temperature. Importantly, the accretion term $f_{\mathrm{{acc},A}}$ has dependence on gas temperature, grain number density, gas number density, and grain size. Grain number density, in turn, increases both with increased pressure, and with decreased gas-to-dust ratio, all else being held constant.

    The desorption rate term $f_{\mathrm{{des},A}}$, on the other hand, depends of grain temperature (assumed equal for gas and dust), the number density of grain surface binding sites (which, for spherical grains of identical sizes, can be calculated from grain number density, and the grain size), and the number density of ice species A per unit volume. This last factor means that, if one assumes an initial abundance of species A in the ice, and set all physical parameters constant such that the initial desorption rate term $f_{\mathrm{des,A}}>f_{\mathrm{acc,A}}$, then $f_{\mathrm{des,A}}$ will decrease over time, as more and more of species A desorbs from the ice phase into the gas phase, thereby decreasing $n_\mathrm{s,A}$. Simultaneously, $f_{\mathrm{acc,A}}$ will increase over time, as the number gas number density increase with the added, desorbed ices. With time, an equilibrium between the gas and ice abundances of species A may be reached, as the desorption and accretion rate terms become similar in size $f_{\mathrm{des,A}}\sim f_{\mathrm{acc,A}}$.

    The above example is a static case, with no changes to the physical conditions over time. If the physical conditions do change over time, or if other initial chemical abundances are assumed, the framework becomes more complicated. In other words, in the chemical kinetics picture, an iceline radius or region is more complex to determine and treat than by using the simpler approach, where the midplane gas temperature is the sole factor for determination.

    \begin{figure}
    \centering
    \includegraphics[scale=0.8]{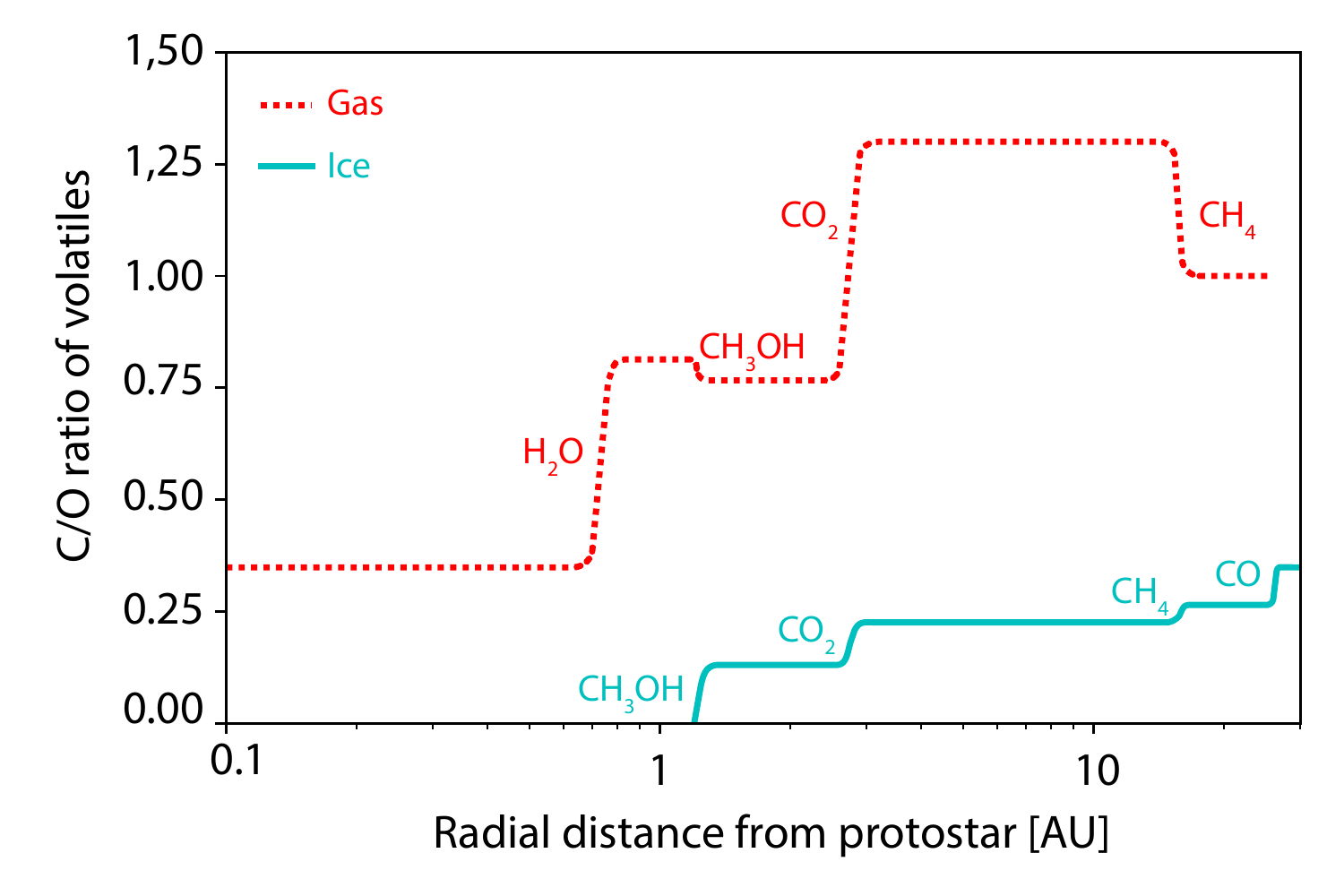}
    \caption{A C/O ratio diagram, accouting for the balance between freeze-out and desorption, leading to iceline regions of gradual instead of abrupt change from gas to ice moving outwards in the disk (see Section \ref{icelinedef}, Point 3). In comparison with Fig. \ref{obergco}, this diagram has \ce{CH3OH} and \ce{CH4} added as species \citep[with binding energies and abundances as described in][Table 1]{eistrup2016}, no carbon or oxygen is assumed to be in silicates or refractory carbon, and the global C/O ratio here is 0.34, whereas it was 0.55 (solar) in Fig. \ref{obergco}. This C/O ratio diagram is an artist's impression, made by the author.}
    \label{eistrupco}
\end{figure}

    \item Definition of an \textbf{ice surface} (snow surface) by including the vertical dimension and physical structure of the disk. The inclusion of the vertical structure of the disk is more realistic, but also more complex.

    Firstly, it is a necessary inclusion if one wants to investigate and model the effects of vertical mixing caused by turbulence, where gas and dust from the midplane is cycled to the upper layers of the disk, and where, in turn, material from the upper layers are cycled down to the midplane. Because the upper layers are warmer, are exposed to higher levels of radiation and ionisation, and have different chemical compositions than the midplane, this vertical mixing can change the chemical composition throughout the vertical dimension. Furthermore, the assumption that $T_{\mathrm{gas}}$=$T_{\mathrm{grain}}$ is likely not realistic when moving away from the midplane, and this may, in turn, results in different chemical interplay between gas and grains.

    Secondly, the vertical dimension is necessary to include if one wants to infer the radius of a given midplane iceline from observations. This is because the emission from disks usually arises in the upper layers of a disk, rather than in the midplane. And if the upper layers have a different physical structure than the midplane, then the distance from the central star where freeze-out and desorption balance each other out tend to be further away from the star than is the case for the midplane iceline radius, because the upper layers are warmer. Following this, instead of a midplane \emph{iceline}, an ice \emph{surface} can be drawn from the upper layers of the disk through the disk midplane, indicating where, at each disk height, species A will transfer from gas to ice. One example of such an inference of disk midplane iceline radius was the work on the iceline radius for CO in the PPD around the young star TW Hya by \citet{qi2013}, and subsequently the analysis of the robustness of this method by \citet{merelhoff2017}, where the concept of a CO snow (ice) surface was outlined, see Fig. \ref{merelco}. As a contrast to this, \citet{bosman2021water} reported that in the PPD around the young star AS 205, the \ce{H2O} ice surface was observationally determined to be at the same radius in the midplane and in upper layers of the disk.

    \begin{figure}
    \centering
    \includegraphics[scale=0.8]{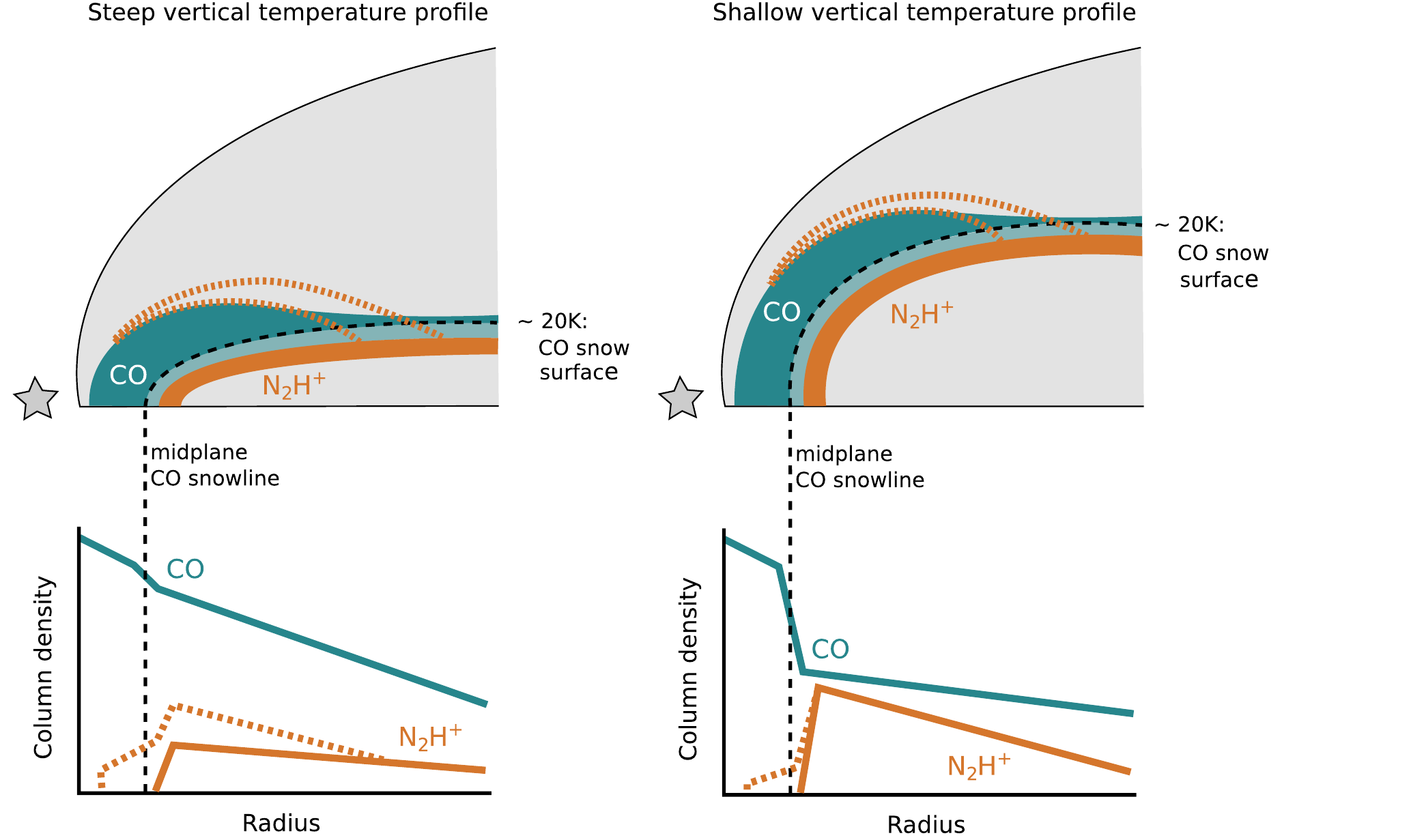}
    \caption{The top two schematics indicate with black dashed profiles the expected locations of the CO snow (ice) surface, as a function of midplane radius and disk height. It is seen that the closer one is to the midplane, the closer this location is to the central star. Reproduced from Fig. 11, \citet{merelhoff2017}. Copyright 2017 Astronomy \& Astrophysics.}
    \label{merelco}
    \end{figure}

\end{enumerate}

In choosing how to define the iceline of species A in a disk midplane, one has to weigh the ease-of-implementation of Point 1, against the more rigorous dynamical, chemical and disk structural treatments of Points 2-4, which ultimately is a matter of how many spatial dimensions one wants to work with (chemical reactions still excluded). As will be laid out in the remainder of this review, in order to further the connection between realistic astrochemistry and realistic planet formation modelling, it is important that the choice of iceline definition is both spelled out and motivated when undertaking research in the field of planet formation chemistry. In addition, it is important to consciously discuss the drawbacks of the simpler iceline definitions, and attempt inclusion of more realistic iceline treatment in the future.

\subsection{Limitations of the classical ``iceline'' picture}

In this section, the classical picture of planet formation chemistry has been outlined, using what can be described as ``iceline chemistry''. This idea was to reduce chemical complexity by only considering two chemical effects, namely freeze-out and desorption, which relate to the two physical phases that molecules are considered to be in in PPD midplane, namely gas and ice. The benefit of this approach was that it allowed insights into the elemental carbon and oxygen (as well as other elements) content of the gaseous and icy material in different parts of the disk midplane, which could then, in turn, be used to estimate the elemental ratios of the (exo)planetary atmospheres that could be formed in different parts of the disk midplane. However, the classical chemical picture does not consider any chemical reactions to take place between chemical species. The chemical composition in the disk midplane is considered inert.

This is a simplification. The last decade have seen efforts being made to assess which types of chemical reactions are likely to take place under PPD midplane physical conditions (including a time scale limitation of around 10Myr), what the rates of these reactions are, and which chemical changes to the midplane composition these reactions might result in. These chemical effects will be outlined in Section \ref{chemreac}.

\section{Including chemical reactivity}
\label{chemreac}

Chemical reactions take place in space, including in planet-forming disks around young stars \citep[see, e.g., reviews by][]{herbst2009review,dishoeck2014faraday,dishoeck2017review}. This reactivity, in turns, needs to be accounted for when modelling the composition of forming planets. Whereas much of the work assuming iceline chemistry simply used sets of abundances that reflected the volatile compositions of interstellar ices, and assumed these abundance sets to be chemically unaltered \emph{en-route} until incorporation into at planet, it is more realistic to assume that chemical reactions take place underway, and that these reactions change the chemical composition of the planet-forming material, from the beginning till the end of planet formation.

There have been different approaches to this. One is to assume that the chemical composition of the disk midplane is in local thermodynamic equilibrium (LTE), and then use this to partition the chemical elements (C, N, O, H, etc.) between molecular carriers. This has the advantage that codes for modelling LTE chemical abundances are widely available, and the modelling is quickly done. However, a basis of such an assumption is that the change from interstellar medium (ISM) ice abundances to LTE abundances happens quickly, which, given the temperatures and pressures characteristic to a PPD midplane, along with the lifetime of the PPD, is unrealistic. Therefore, PPD midplane chemistry should not be modelled using LTE abundances (except very close to the star, inside the \ce{H2O} iceline, where the temperature and number density is high, and all volatile species are in the gas-phase).

\subsection{Chemical kinetics}
\begin{figure}
    \centering
    \includegraphics[scale=0.7]{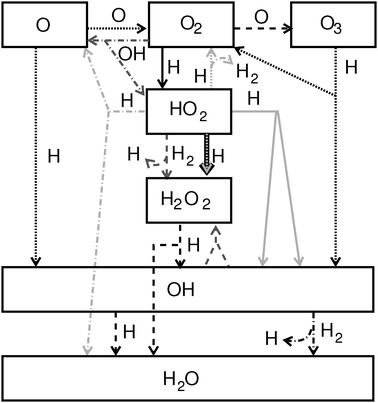}
    \caption{Reaction network for grain-surface reactions involving O and H. In such a network, a chemical reaction is represented by an arrow. Each arrow has a species next to it (a reactant), and the arrow is pointing from one species (the other reactant) to another species (the product). Reproduced from \citet{lamberts2013}. Copyright 2017 Physical Chemistry Chemical Physics.}
    \label{reacnetwork}
\end{figure}

A more realistic approach is using chemical kinetics, albeit also more complex and computationally expensive than LTE modelling. The basis of chemical kinetics is three-fold:
\begin{itemize}
    \item Every chemical species has a kinetic binding energy $E_\mathrm{bind}$ associated with it, which influences the partitioning of the species between the gas and ice phases under different physical conditions.
    \item Every chemical reaction, which include reactants and products, has a rate, which is the efficiency of this reaction (how many reactant molecules are destructed, and how many product molecules are produced per second) at a given temperature. This rate can be calculated, and also considers factors such as particle density (pressure), sizes of grains, level of ionisation (\citep[which is expected to vary from disk to disk, and to have notable effects on disk midplane chemical evolution, see][]{cleeves13crex,eistrup2016,schwarz2019co,bosman2018,kuffmeier2020ion,aikawa2021mapsion}), radiation levels, and kinetic binding energies of reactants.
    \item Given a initial set of chemical species, and their initial abundances (usually expressed as a number density relative to H), a chemical network of reactions with rates, and reactants and products with kinetic binding energies, can be set to \emph{evolve} in time, by calculating the rates of each reaction, and change the abundances of each chemical species accordingly, for a preset time step forward. An example of such a network can be seen in Fig. \ref{reacnetwork}, for grain-surface reactions involving O and H. These computed new abundances of each species are then saved and used as starting abundances for the next chemical evolution time step forward. This is repeated for as long time as one assumes chemical evolution to take place over.
\end{itemize}

In other words, chemical kinetics does not compute a set of equilibrium chemical abundances which one can assume as input for planet formation chemistry (although, given a long enough chemical evolution time, the abundances of each chemical species considered can converge as time step-to-time step changes decrease, one can end up with a set of \emph{steady state} abundances, which will then not change considerably if longer time for chemical is assumed). Instead, chemical kinetics assumes an initial set of chemical abundances, and then computes the time-dependent changes to these abundances. By making assumptions about where is a disk midplane this chemical evolution takes place, and for how long time it takes place before material is incorporated into planet, one can estimate a more realistic chemical composition of the material forming planets than in the case of a chemically inert pre-set set of abundances.

A chemical network for use in chemical kinetics calculations can, in principle, account for various chemical states, phases and reaction types, some of which are outlined and discussed in-depth in \citet{cuppen2017review}. The ability of chemical kinetics networks to include both gas-phase chemical reactions, and reactions between ices on the surfaces of grains, while simultaneously accounting for the interactions between the gas and the ice phases (through freeze-out and desorption) makes chemical kinetics very useful for modelling chemical evolution in PPD midplanes, where physical conditions vary and accounting for both gas and ice phases is necessary. However, the usefulness of a chemical kinetics reaction network depends on how well-constrained the reaction rates and binding energies are, and on the degree to which the chemical effects at work in PPD midplanes are actually represented by by reaction types in the network.

Binding energies and reactions rates are estimated both through quantum chemical calculations and through experimental studies in astrochemical laboratories \citep[see, e.g., reviews by][]{cooke2019review,potapov2021review}. However, more results and data have been produced for simpler species, like CO and \ce{H2O} than for larger molecules, including many complex organic molecules (COMs). COMs are molecules with both C and O-atoms, and which consist of six of more atoms. \ce{CH3OH} (methanol) is considered the simplest COM. Likewise, binding energies are also better estimated for simpler species, than they are for complex species. Lastly, it is generally more challenging from a laboratory astrochemistry point-of-view to experimentally estimate reaction rates for ice-phase reactions on the grain surfaces than it is to estimate pas-phase reaction rates, not least because assumptions need to be made about the reactivity and volume of the surface layers of grains, respectively, the bulk of ice underneath these layers. Nonetheless, ice-phase chemistry possibly plays an important role in chemical evolution in PPD midplanes, and hence it is important to account for these reaction types when modelling chemical evolution, even if the reaction types included in a network is not a accurate representation of reality, and the rates of these reactions are less well-constrained.

Lastly, it is noted that detailed microscopic simulations of grain-surface chemistry is also being performed, using Kinetic Monte Carlo (KMC) methods, see \citet{cuppen2013,willis2017mcmc}. Results from such simulations are an important addition to the use of rate equations in chemical kinetics, as depicted in Fig. 1 of \citet{cuppen2013}. KMC-methods provide detailed insights into the behavior of different chemical species under different physical conditions and for different types of grain surfaces. However, these simulation are computationally expensive. Therefore, the statistical generalisations of KMC simulations which are used for rate equation in chemical kinetics, are often considered as the best of both worlds, achieving some degree of chemical complexity from KMC-methods, whilst being computationally doable for a larger number of chemical species and reactions (by virtue of the rate equation approach). The trade-off is that much microscopic detail (which is accounted for in KMC simulations) is neglected. KMC methods will not be discussed further in this review.

\begin{figure}
    \centering
    \includegraphics[scale=0.8]{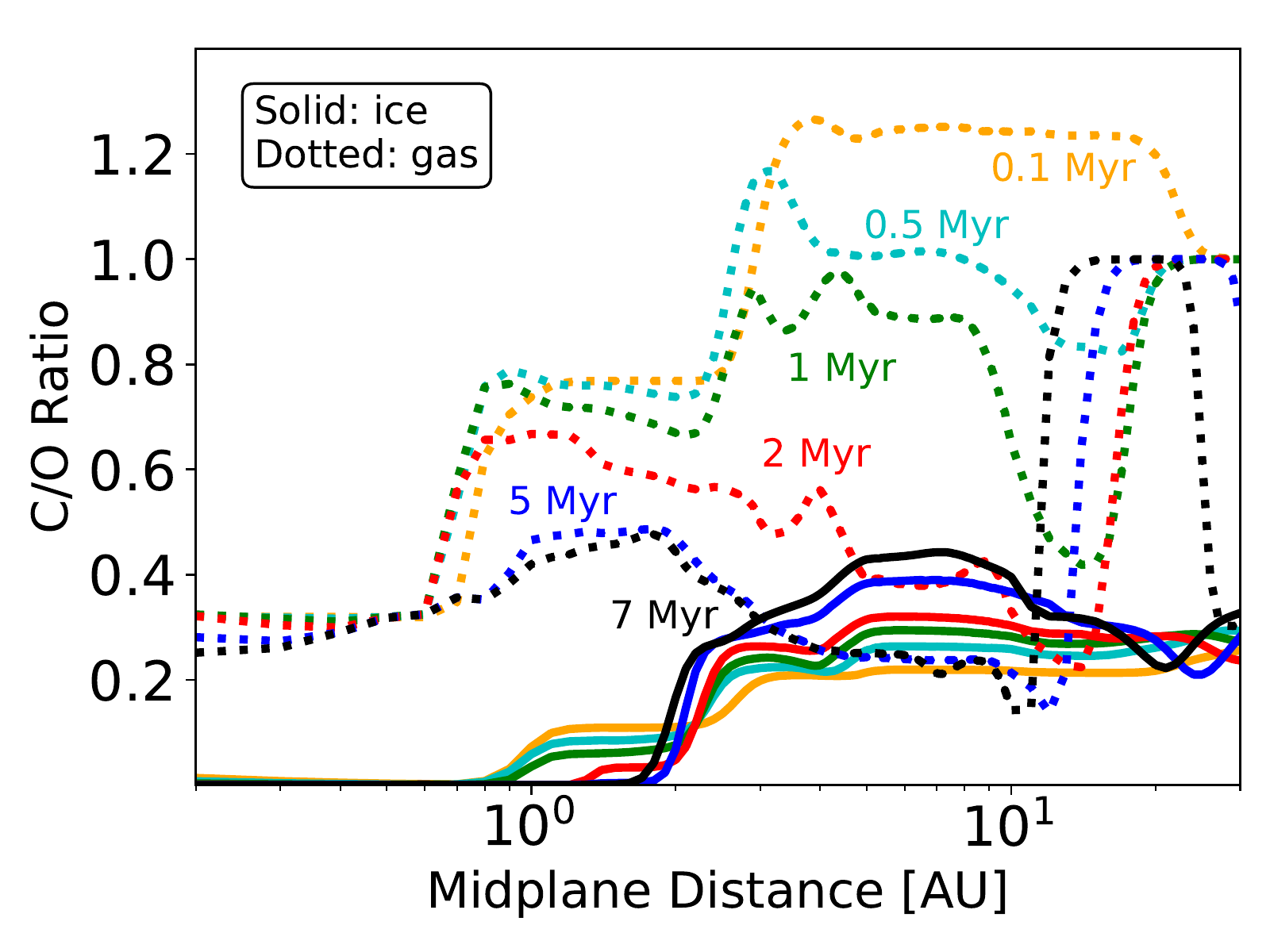}
    \caption{Adapted from Fig. Fig. 9.a in \citet{eistrup2018}. C/O ratios for volatile species in gas and ice, as a function of disk midplane radius, and as a function of evolution time (from 0.1-7 Myr). Underlying physical disk structure is evolving over time. No dynamical effects like drift of solids are included. The stellar (global) C/O ratio here is 0.34. Reproduced from Fig. 9.a, \citet{eistrup2018}. Copyright 2018 Astronomy \& Astrophysics.}
    \label{eistrupcoevol}
\end{figure}

Fig. \ref{eistrupcoevol} \citep[adapted from Fig. 9.a in][]{eistrup2018}) shows the effects that chemical evolution can have on the C/O ratios of gas and ice in a disk midplane, by adding the time dimension. The disk physical structure is evolving, so icelines are seen to shift over time. Axes are similar to those in \ref{eistrupco}: C/O ratio (linear) and disk midplane radius (log), and the yellow profiles are comparable to the profiles in Fig. \ref{eistrupco}.

\subsection{Choosing the degree of chemical complexity}
\label{complexity}

It may be tempting to jump head-first into applying the most expansive and comprehensive chemical network available, in order to model PPD midplane chemistry as realistically as possible. However, this may not be the optimal approach, since, indeed, many of the reaction rates are not well-constrained, and there is uncertainty about which reaction types are important in a given physical environment, and which are not. The results of such modelling, in turn, need to be compared to and constrained by observed chemical compositions of PPDs midplanes, which are still scarce. One effort into probing PPD midplane abundances was carried out by \citet{zhang2017}.

A reasonable alternative is to start with a simple chemical network (with the more well-constrained and understood reaction types), and then expand this network to include more and more chemical complexity. \citet{fogel2011} used a chemical kinetics network with gas-phase reactions, freeze-out and desorption, and hydrogenation on grain surfaces as the only chemical effect in the ices. The same network and code was used by \citet{cridland2016,cridland2017} in order to implement chemical evolution into a planet formation model.

\begin{figure}
    \centering
    \includegraphics[scale=0.4]{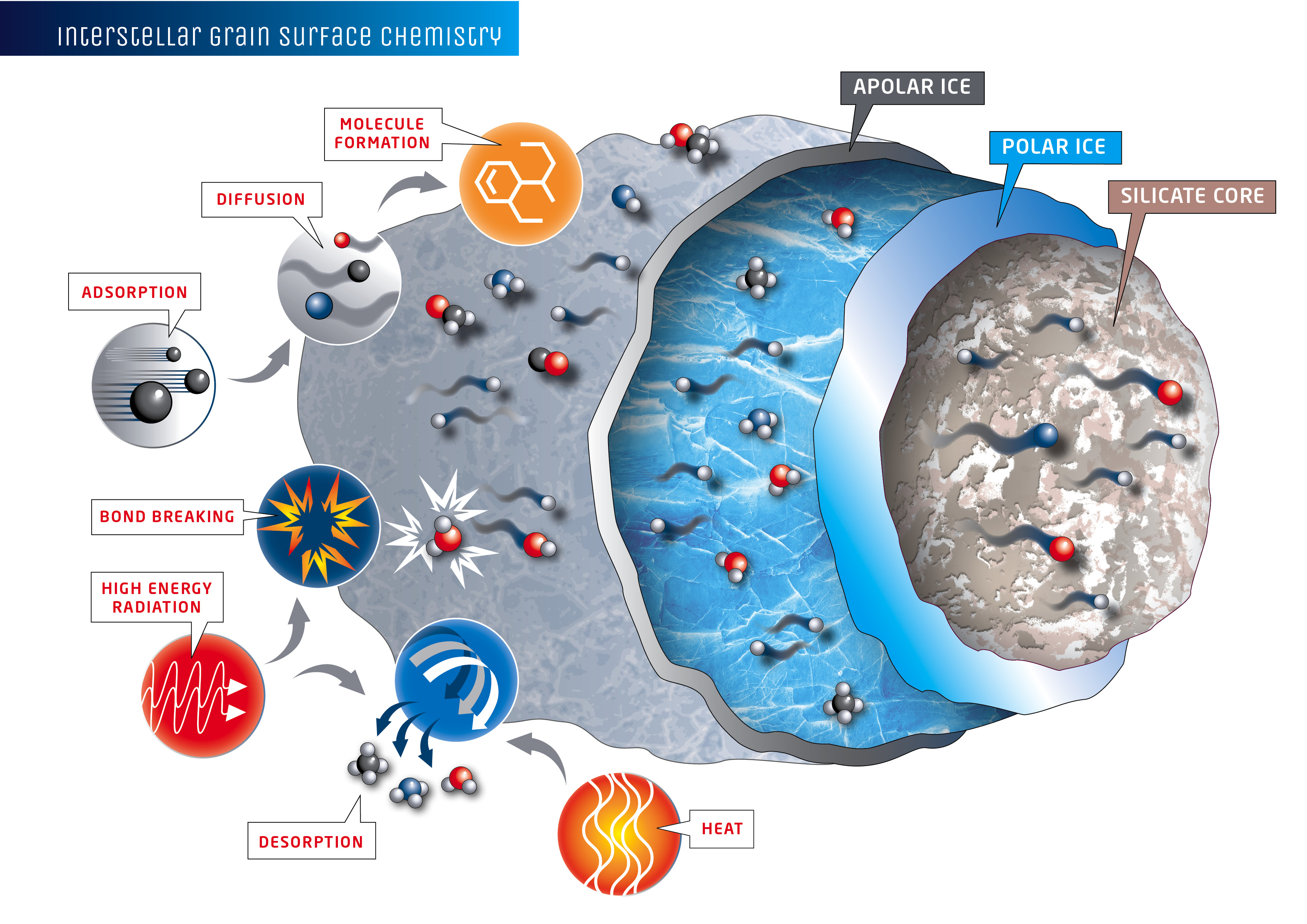}
    \caption{Chemical effects related to dust grains. The top layer of the grain (denoted ``apolar ice'') is usually referred to as the grain surface, and the middle layer (denoted ``polar ice'') is usually referred to as the bulk of the ice (when utilised three-phase chemical kinetics codes). Reproduced from: Leibundgut, Drozdovskaya, Wampfler \& Ligterink. Open Access.}
    \label{graincartoon}
\end{figure}

Hydrogenation is the process when ices on the grain surfaces react with hydrogen atoms. One example is the process whereby CO ice can react to form first HCO, then \ce{H2CO} (formaldehyde) and finally \ce{CH3OH}. Thereby, hydrogenation is one proposed way whereby chemical reactions on icy grains can lead from simple to complex chemistry. Fig. \ref{graincartoon} provides an impression of the chemical effects relating to dust grains.

However, hydrogenation is only of likely type of reaction taking place on grain surfaces. Another is general two-body grain-surface reactions, where neither reactant is hydrogen. Hydrogenation reactions are different because hydrogen finds reaction partners on the grain surface so fast that these reactions are effectively instantaneous, whereas two-body reactions not involving hydrogen, are not. One such reaction is \ce{iCO + iOH -> iCO2 + iH} (where ``i'' denotes ice-phase), which produces \ce{CO2}. The reactant OH can be a product of a reaction dissociating \ce{H2O} into OH and O. In the classical iceline chemical picture the reaction above producing \ce{CO2} from CO and OH will only proceed when both reactants are frozen out, and given the low binding energy for CO (855K), this will only happen in the outer disk midplanes below $\sim$ 20K. However, in the chemical kinetics picture, where freeze-out and desorption are mechanisms at play under all physical conditions, some CO molecules will collide with grains at all radii in the disk. For very high temperatures, the CO adsorption time when colliding is of the order of a vibrational period, so the CO will not actually stick to the grain. However, at lower temperatures, although still above the thermal CO iceline temperature, CO can stick to the grain, long enough to be available for grain surface reactions. If then, in turn, the reaction rate for producing \ce{CO2} from CO and OH is higher than the rate for desorbing CO off the grain surfaces (and this under physical condition that would place CO in the gas-phase, thus, e.g., for midplane temperatures above 20K), then CO sticking to the grain will react with OH before the CO has time to desorb, and chemical kinetics thereby predicts that some CO will be taken out of the gas-phase and will lead to an increase in the \ce{CO2} ice abundance on the grain surfaces (that is, when the midplane temperature is low enough for \ce{CO2} to remain in the ice-phase, which, given $E_\mathrm{bind}$(\ce{CO2})=2990K happens below $\sim$ 60K for PPD midplane conditions).

This effect has been modelled by \citet{eistrup2016,schwarz2016,bosman2018}, and used to explain the low abundance of CO gas observed towards multiple PPDs \citep[see, e.g.,][]{zhang2017,schwarz2019co,alarcon2021mapsco}. This effect also leads to a change in the overall carbon-to-oxygen ratio picture from Fig. \ref{eistrupco}, as some elemental C and O carried in CO, which, between the \ce{CO2} and CO iceline in the classical picture, were in the gas-phase, are now processed into the ice-phase in this radial midplane region, thereby contributing to an alteration of the partitioning of C and O in gas and ice, as a function of radius. Another outcome of chemical kinetics with grain-surface chemistry that also alters the C/O ratio partitioning in gas and ice is when two ice species with relatively high binding energies react to form a product with a lower binding energy. If the physical conditions allow for the reactants to mostly be in the ice-phase, and the product(s) to mostly be in the gas-phase, then such a chemical reaction acts to effectively transform the constituent elements (e.g., C and O). of the reactants from the ice-phase into the gas-phase. This, once again, can alter the partitioning of C and O between the gas and ice-phases, as a function of radius.

\begin{figure}
    \centering
    \includegraphics[scale=0.7]{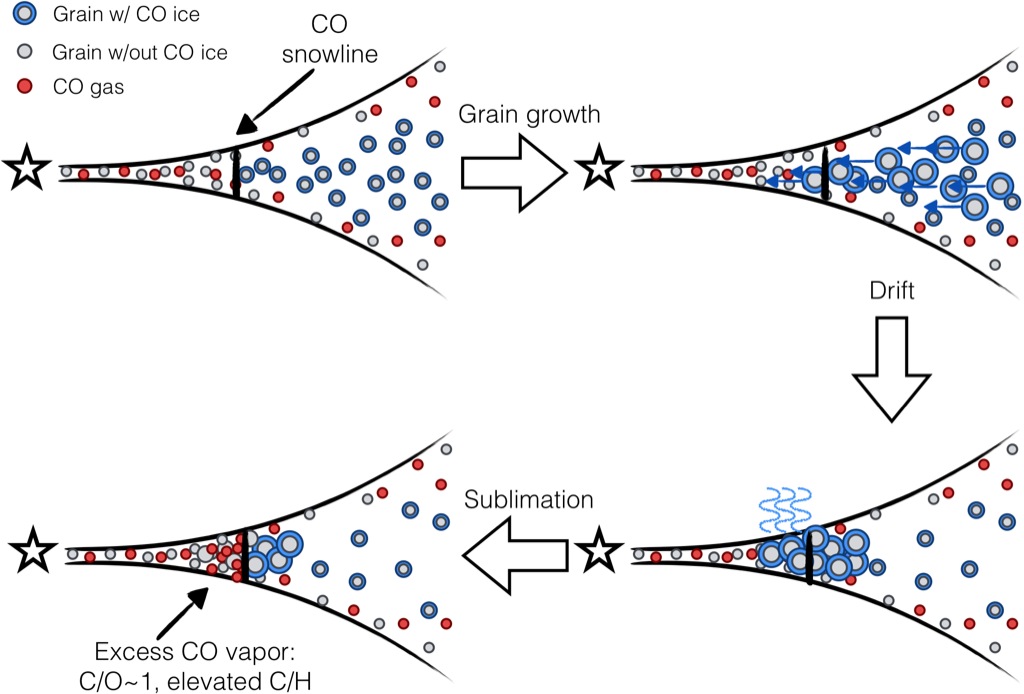}
    \caption{Cartoon view of the region around the CO snowline. Initially the gas and grains are co-located and the CO gas abundance with respect to \ce{H2} is determined by a balance between freeze-out and desorption. As grains coagulate and grow to form pebbles, they begin to drift. CO-ice-covered pebbles that drift across the CO snowline will sublimate, locally altering the gas-phase composition inside the CO iceline. Reproduced from \citet{oberg2016}. Copyright 2016 Astrophysical Journal.}
    \label{obergdrift}
\end{figure}

Another effect which chemical evolution with chemical kinetics can have on the C/O ratio in the gas was found in \citet{eistrup2018}: ice species with higher binding energies, such as \ce{H2O}, \ce{CO2} and \ce{OH} ($E_{\mathrm{bind}}$=5770K, 2990K and 2850K, respectively), are dissociated, and the resulting O atoms react to form \ce{O2} ice. However, \ce{O2} has a significantly lower binding energy of $E_{\mathrm{bind}}$=1000K, which means that \ce{O2} is mainly in the gas-phase for disk midplane tempatures above 30K. These reaction pathways therefore lead to elemental oxygen being processed from the ice into the gas, for temperatures between the freeze-out temperatures of \ce{CO2} and OH at $\sim$60K and the freeze-out temperature of \ce{O2} at $\sim$30K.

The choice of chemical complexity for inclusion of chemical evolution into planet formation models ultimately depends on several factors:
\begin{itemize}
    \item \textbf{The physical conditions} in the region of the disk midplane under consideration. If the inner, hotter midplane in considered, then grain-surface chemistry will have a relatively minor effect compared to gas-phase chemistry, simply because more species are in the gas than is the ice. One cautious note, however, is that solid bodies drifting dynamically from the outer disk to the inner (see, e.g., \citet{lambrechts2012} and Fig. \ref{obergdrift}) will bring icy material with them which have undergone ice chemical processing under conditions different from those in the inner disk. Therefore, when these bodies reach the inner disk on a timescale of $10^5-10^6$ year (accounting for both grain growth and chemical evolution in the other disk, as well as the actual drift), then different ices carried on them maybe desorb and mix with gas which was in the inner disk throughout the evolution, which alters the local gas composition and elemental ratios.

    If the outer, colder disk midplane is considered (say, outside the CO iceline), then almost all chemical processing will take place in the ices. Hence, under these conditions, a comprehensive network of reaction in the ices is useful. There are, in turn, special conditions that may affect which reactions types are necessary for the ices. If the local temperature is \textless 10K, then there will generally not be sufficient thermal energy for molecular, and some atomic, species to move from binding site to binding site (in chemical kinetics, a certain amount of sites/spots/places is assumed per grain surface area for individual species to stick to when they freeze out from gas to ice, and a physical energy barrier is assumed to be present between each site, which a mobile species needs to ``hop'' over or tunnel through, in order to move) around on the grain surface with the possibility of encountering a species to react with at some site. In this case, only hydrogenation reactions are considered relevant, since hydrogen (atomic and molecular) is able to quantum tunnel through such barriers, which are generally assumed to have a width of 1-2 \AA, see \citet{cuppen2017review}.

    If, however, the temperature is higher, then two-body reactions in the ices play a role as well, and at temperatures above 20 K, these reactions will likely dominate over hydrogenation reactions, since also hydrogen, with binding energies of either 600 or 430K for atomic, respectively, molecular hydrogen, is mostly in the gas-phase for midplane temperatures above 15K. Some chemical effects on ice abundances caused by different choices of grain-surface barriers widths, and of the mobility of larger surface species (or ability for each species to diffuse around the grain surface, parameterised as the fraction of diffusion-to-binding energy $\frac{E_{\mathrm{diff}}}{E_{\mathrm{bind}}}$) which need to thermally hop over barriers in order to find each other and react chemically, was explored in \citet{eistrup2019o2}.

    In-between these two temperatures extremes, the interactions effects between the gas-phase and the grain-surfaces may be important, along with gas-phase reactions and ice reactions themselves. Therefore, a chemical kinetics network is very useful, and should include freeze-out and desorption terms, as well as two-body reactions in the ices.

    \item \textbf{Timescales}. Gas-phase reactions, along with freeze-out and desorption, are relatively short-term effects (\textless 10$^4$ yrs), whereas reactions in the ice-phase are relatively longer term ($> 10^5$ yrs). Therefore, one needs to consider which timescales each element of planet formation is assumed to take place over, in order to estimate the type of chemical evolution than is reasonable to account for in the disk, prior to planet formation commencing

    On very long timescale ($>$1Gyr), there are additional chemical effects that may influence chemical evolution, not just on the surfaces of solid bodies, but also inside them. \citet{garrod2019comets} modelled chemical evolution inside comets on these timescales, and found that energetic radiation and particles could penetrate comets enough to cause chemical changes down to $\sim 10 $m. This, in turn, may have an impact on later-stage chemical evolution for planets that are impacted by comets, Myrs or Gyrs after they initially formed.

\end{itemize}

\section{Tools and codes for chemical kinetics}
\label{tools}
Some chemical kinetics codes suited for modelling disk midplane chemical evolution have already been mentioned. However, below is a short outline of some codes to try out and test for one's own purposes:

\begin{itemize}
    \item Chemical kinetics code\footnote{Please refer to L. Ilsedore Cleeves at University of Virginia, United States. Email: lic3f@virginia.edu} by \citet{fogel2011}, which is based on the Ohio State University (OSU) Astrophysical Chemistry Group gas-phase model network from 2008 March, see \citet{smith2004}. This network includes gas-phase reactions, gas-grain interactions, and some hydrogenation reactions in the ice. This code is written in Fortran.
    \item The \textsc{Nautilus} code \footnote{Please refer to Valentine Wakelam at Laboratoire d’Astrophysique de Bordeaux, France, and http://perso.astrophy.u-bordeaux.fr/$\sim$vwakelam/Nautilus.html. Email: valentine.wakelam@u-bordeaux.fr} by \citet{wakelam2012}, and also used by \citet{taquet2016}. This network is based on the Kinetic Database for Astrochemistry (KIDA) chemical network, and includes gas-phase reactions, gas-grain interactions, and grain-surface reactions. Alongside the model used by \citet{furuya2015}, these chemical codes also consider the bulk of the grains to be a chemical phase separate from the gas-phase and grain surface, in which a chemical species can reside. These chemical kinetics models are therefore sometimes referred to as multi-phase or three-phase models (see also Fig. \ref{graincartoon}), as opposed to models including only the gas-phase and grain surfaces, which are referred to as two-phase models.
    \item The \textsc{Alchemic} code\footnote{Please refer to Dmitri A. Sememov at Max Planck Institute for Astronomy, Heidelberg, Germany. Email: semenov@mpia.de} by \citet{semenov2010}, and the \textsc{Walsh} code\footnote{Please refer to Catherine Walsh at University of Leeds, United Kingdom. Email: C.Walsh1@leeds.ac.uk} from \citet{walsh2015} which are both two-phase models that include gas-phase reactions, gas-grain interactions and comprehensive grain-surface chemistry. Both these codes are written in Fortran.
    \item The \textsc{KROME} package\footnote{Please refer to http://kromepackage.org/index.html} by \citet{grassi2014}. This package contains a code wrapper written in Python, which can be used to generate the Fortran subroutines (which is the starting point for the codes mentioned above) that carry out the chemical modelling. The \textsc{KROME} package can handle any chemical network provided by the user (including gas-phase reactions, gas-grain interactions, grain surface reactions, and reactions in the bulk of the grains), and as such this package eases the implementation of chemical kinetics modelling into any numerical code (e.g. for planet formation).
\end{itemize}

\section{Connecting chemistry and physics for modelling planet formation}

Until this point, this review has focused on what chemistry in disk midplanes is, how it can be modelled, and which types of evolution it might lead to. It has been highlighted that chemical evolution may have profound and important implications for the process of planet formation, and for predicting the chemical compositions of planets.

This section will focus on how chemistry has been implemented into models of planet formation, and, vice versa, how planet formation mechanisms have been implemented into chemical evolution models. The section will include a discussion of suggested best practices on how to model planet formation and chemical alongside each other.

\subsection{Implementing chemistry into physical simulations}
\label{connect1}

Given the advances made in understanding and modelling the physical processes involved in planet formation, it is perhaps understandable that big efforts have been put into implementing chemistry into the these models. These implementations vary in the complexity of their treatments of chemistry. The zeroth-order approach, as previously mentioned, is the iceline chemistry framework. In this framework, an initial chemical composition is assumed, for both volatile species (ices) and for refractory components (minerals).

One common way to do this is to assume inherited molecular ice abundances from the local ISM \citep[see, e.g.][]{oberg2011co,alibert13}, hence with no chemical evolution from molecular cloud-to-protoplanetary disk midplane stage. The inheritance of insterstellar ices in a PPD was modelled by \citet{bergner2021inh}. Similarly, initial abundances can also be based on solar system cometary abundances, as in the study by \citet{ilee2017}. However, cometary abundances may not be representative of the actual ice abundances prior to planet formation, given the possible chemical evolution in cometary ices (see \citet{garrod2019comets}), and the variation in volatile abundances observed across comets studied by \citet{leroy2015}. Another way is to assume the elemental ratios (C/H, O/H, N/H, etc.) from host star abundances, and then use LTE to compute the partitioning of the elements into refractory and volatile species in the disk midplane \citep[e.g.][]{madhu2014,bitschbattistini2020, bitschsavvidou2021}.

Once a set of initial abundances has been assumed, iceline chemistry is applied. In this context, as is demonstrated in, e.g., \citet{madhu2014,alidib2014, bitschbattistini2020}, it becomes relatively straight forward to determine the composition of gas and ice at any point in the disk midplane from which a forming planet may accrete material, and hence it is easy to derive the resulting chemical composition of such a planet. However, this framework is absent a treatment of chemical \emph{evolution} during the disk evolution and planet formation process.

The planet formation models by \citet{turrini2021} used the disk midplane abundances resulting from 1 Myr of chemical evolution run with a chemical kinetics code in \citet{eistrup2016}, but assumes these evolved abundances to be inert at the start of their planet formation models. This means that the chemical abundances in \citet{turrini2021} were constant (and followed iceline chemistry), but they were based on the results of disk midplane chemical evolution models, rather than on ISM or cometary ice abundances (although the specific chemical setup behind the utilised results from \citet{eistrup2016}, which were assuming chemical evolution starting with inherited ice abundances from the ISM and a low degree of ionisation, actually resulted in abundances that, after 1 Myr of evolution, were very similar to the initial ISM ice abundances). An interesting results from \citet{turrini2021} is that potential exoplanet formation history degeneracies caused by the exclusive use of C/O ratios could potentially be broken by adopting ratios involving N, like C/N or N/O ratios.

A more complex treatment of chemistry for planet formation has been applied in studies such as \citet{cridland2016,cridland2017}, where the chemical kinetics code from \citet{fogel2011} was used (see Sec. \ref{tools}), thus with gas-phase chemistry and balancing of freeze-out and desorption defining icelines, but absent ice chemistry beyond hydrogenation reactions. In \citet{cridland2019comainseq} this was used to derive a ``main sequence'' for C/O in hot exoplanetary atmospheres, enabling a differentiation between the amount of accretion of ices from solid bodies that has gone into forming an atmosphere. \citet{cridland2019_carbon} utilised results from chemical evolution models by \citet{eistrup2018} as the astrochemical basis of the planet formation models. However, here, the chemical evolution (including complex ice chemistry) had already been run under disk midplane conditions, and was thus not running simultaneously with the planet formation models.

The \textsc{KROME} package \citep{grassi2014} was used by \citet{boothilee2019} to implement a treatment of chemical evolution with sophisticated disk dynamics, including grain growth and drift. The chemical network included gas-phase reactions, gas-grain interactions and hydrogenation reactions for ices, following the previous work of \citet{ilee2017}. This work ran physical and chemical evolution simultaneously, and showcased the power and versatility of a chemical kinetics tool like the \textsc{KROME}, for studying chemical effects combined with disk midplane solids drifting independently of the gas, and its overall effect on elemental ratios such a C/O. They found that pebble drift enhances the abundances of volatile species in the inner disk, as these species were carried in ices from the outer disk.

As a last example in this non-exhaustive list of work within this field, \citet{krijtbosman2020} applied sophisticated physical models with both vertical mixing of material between the upper disk layers and the midplane, radial drift of solid material in the midplane, and grain growth, and ran this evolution simultaneously with chemical evolution. The chemical kinetics network was reduced in size compared to other networks (included \ce{H2O}, \ce{CO}, \ce{CO2}, \ce{CH4}, and \ce{CH3OH}), but this reduction was used to gain the computational ability to account for complex ice reactions (including two-body reaction) in addition to gas-phase chemistry, gas-grain interactions and hydrogenation reaction, all whilst simultaneously modelling the physical evolution. This paper demonstrated the power in choosing to ignore more complex (and less abundant) ice species in order to focus of complex grain-surface reaction types that involve simpler and generally more abundant ices. They showed that combining chemical evolution with ice sequestration in the disk midplane can result in 100$\times$ lower CO abundance in the outer disk warm molecular layer. When pebble drift was included, the CO gas abundance was enhanced inside the CO iceline, and they pointed out that the removal of solids an CO from the outer disk complicates the use of CO emission as bulk disk mass tracer.

\subsection{Implementing physics into chemical simulations}
\label{connect2}
This section will outline some of the development that chemical kinetics codes have undergone over time, as the field has attempted to implement a more realistic treatment of the physical conditions in space that chemical codes are intended to model chemical evolution for. These codes feature sophisticated chemical reactions and large amounts of chemical species, but the codes have, by design, been limited to treating more simplified physical conditions and effects than was considered in the previous section. One common simplifying assumption is that grains are all spherical and all the same size.

The codes trace back to, at least, \citet{caselli1993,caselli1998rates,aikawa1996,aikawa1997,aikawa1999accre}, with much work having been done since then to improve estimates on binding energies and reaction rate coefficients \citep[based on quantum chemical calculations and laboratory astrochemical experiments, from, e.g.,][respectively]{garrod2007,cuppen2017review}, add new reaction types, and test these chemical networks under different space-like physical environments and over various timescales for chemical evolution.

The work by \citet{walsh2015} utilised the aforementioned \textsc{Walsh} code and a chemical network based on the UMIST Database for Astrochemistry \citep[\textsc{Rate12}][]{mcelroy13} to test how different host star luminosity regimes affected chemical evolution in a protoplanetary disks, considering both radius and disk height. This network included gas-phase reactions, gas-grain interactions and grain-surface reactions (hydrogenation and two-body).

The same network and code was used for modelling chemical evolution in disk midplanes, specifically, by \citet{eistrup2016}. The \textsc{Walsh} code was then, in \citet{eistrup2018}, used to explore the effects of changing physical conditions, such as temperature, density and ionisation, for each chemical evolution time step, in order to mimic a protoplanetary disk midplane that cools as it physically evolves over time. Both these studies found that chemical reaction in gas and in ice can cause species with high binding energies to be processed into species with low binding energies (\ce{H2O} processed into \ce{O2} on grains), and vice versa (CO+OH processed into \ce{CO2} on the grains). These processes can collectively change whether C, O and N is primarily carried in gas or ice species, under given physical conditions. This, in turn, can evolve the classic C/O ratio as a function of radius, over time.

More recent studies have added additional insight into the effect that more realistic physical treatments have on chemical modelling. \citet{gavino2021} modelled chemical evolution as a function of PPD radius and height utilising the three-phase \textsc{NAUTILUS} code \citep{ruaud2016}, and with different assumptions for grain sizes. Grain sizes matter to chemical evolution, as few larger (spherical) grains bodies have a smaller surface areas than many smaller (spherical) grains, assuming the total grain mass to be the same in the two case. A smaller total surface area reduces the availability of grain-surface sites where chemical reactions can take place for ices, and also effectively reduces the probability of gas species encountering a dust grain surface, to freeze out onto, thereby reducing reducing the freeze-out term in Eq. \ref{acc}. Assuming different grain sizes may therefore lead to important new behavior for chemical evolution during planet formation.

A treatment of vertical settling of dust grains from the upper layers of the disk towards the disk midplane was implemented into a chemical kinetics model by \citet{clepper2022graingrowth}. They modelled chemical evolution of ices on grains and pebbles in a midplane, accounting for the effect of turbulent mixing with the upper layers of the disk. They only utilised one model for grain growth. They found CO gas to be depleted from the surface layers of a disk by 1-2 orders of magnitude (similar to the findings of \citet{krijtbosman2020}), assuming an $\alpha$-turbulence with $\alpha=10^{-3}$. This is achieved by a combination of ice sequestration and decreasing UV opacity, both driven by pebble growth, they conclude. Their models are, however, not able to explain C/O ratios of 1.5--2, which implies that other effects in addition to ice sequestration are needed in order to explain higher-than-unity C/O ratio values in disks.

The \textsc{Walsh} code was adopted by \citet{cevallossoto2022}. They accounted for pebbles drifting through new material (gas and dust), as they moved through the disk. For their models with stellar accretion rates of $\dot{m}=10^{-8}M_{\odot}$/yr, they find that only pebbles drifting from radii $<10$AU make it to their inner disk region within 100kyr. For lower stellar accretion rates of $\dot{m}=10^{-9}M_{\odot}$/yr, their pebble drift speeds are faster (drift times shorter), and so pebbles from the outer disk ($>10$AU) can influence their inner disk on at 100kyr timescale. Their work predicts that pebble drift may transport a significant amount of volatiles from the outer disk to the inner disk, with the results that the outer disk loses volatile ices, and the gas abundances around volatile icelines in the inner disk increase.

Lastly, \citet{eistrupkrijtcleeves2022} modelled the effect that grain growth has on the efficiency of chemical reactions on grain surfaces, at fixed radii without any drift. Chemical evolution was modelled with both both evolving grain sizes (which changed for each time step in the chemical evolution), and for different constant grain sizes (from $R$=0.1$\mu$m up to $R$=1mm). It was found that using a constant grain size throughout chemical evolution may be an appropriate simplification, instead of changing the grain sizes for each time step, as long as the chosen constant grain size is in agreement with models for grain growth (thus larger than 0.1$\mu$m). The drift aspect for planet-forming pebbles was then added by \citet{eistruphenning2022}, in which the chemical evolution of ices on drifting pebbles was modelled, assuming different drift timescales. Here it was found that the chemical changes in the ices on drifting pebbles may be negligible, if the drift time scale is short enough (a few times $\sim10^{4}$ yr to drift from 200AU to 1AU). However, for longer drift time scales ($\sim10^{5}-10^{6}$ yr for the same drifting distance), chemical evolution in the pebble ices during drift need to be taken into account.

\subsection{From disk midplane to exoplanet atmosphere \emph{and back}}

The outcome of disk midplane evolution and planet formation is, of course, planets and exoplanets. The research field of exoplanets has emerged and grown dramatically since the first detections of exoplanets by \citet{wolszczan1992,mayor1995}. Exoplanet atmospheres are now being observed and characterised regularly, with very interesting insights into their compositions following \citep{dekok2013,brogi2013,kreidberg2014water43,tsiaras2019,gravity2020betapic,molliere2020}.

The physical conditions in exoplanet atmosphere are such that the chemical elements in it are likely in LTE, and therefore the molecular composition of the atmosphere does not necessarily represent that of the material that formed it. The constituent elements of the atmosphere-forming material, however, are likely conserved (not accounting here for possible erosion of material from the surface of the solid core of an exoplanet into its atmospheres, nor the sequestration of atmospheric material onto the surface of the core). This means that the C/O/N/S/H ratios of the atmosphere-forming material combined should reflect that of the composition of the atmospheric chemical make-up.

By characterising exoplanet atmospheres observationally, and constraining their elemental ratios, one can, in principle, derive information about how the atmosphere formed in the natal PPD midplane. Such inferences are, though, degenerate, in that different combinations of the time of atmospheric formation, different radial locations for accretion of the atmosphere, and different ratio of gas and solids body accretion may result in similar elemental ratios in the resulting atmospheres. However, modelling these parameter spaces does hold a key to connecting planet formation in disk midplanes to exoplanet atmospheric compositions. Atmospheric formation for Hot Jupiter exoplanet atmospheres from different assumptions for disk midplane chemistry was modelled by \citet{notsu_eistrup2020}, and, for the modelling setups that were tested, some conclusions could be drawn about exoplanet atmospheric formation histories, depending on which C/O ratio the atmosphere was assumed to have.

One example of how observed atmospheric abundances are used to infer formation histories of (exo)planets was published by \citet{obergwordsworth2019,bosman2019jupiter}. These two paper both utilised the higher-than-solar atmospheric abundance of elemental nitrogen in Jupiter's atmosphere to infer that it must had formed from accretion of drifting \ce{N2} ice-covered solid bodies (pebbles) near or outside of the \ce{N2} iceline in the pre-solar nebula.

Recently, \citet{molliere2022} undertook a major step towards enabling a statistical connection between observed and characterised exoplanet atmospheres, and planet formation histories. They established a Bayesian framework upon which an exoplanet with characterised atmosphere can be statistically coupled to its most likely formation region in a protoplanetary disk midplane, dependent on how the chemical evolution, and thereby the evolution of the elemental ratios in gas and ice in the midplane, changes over time to possibly fit with the observed exoplanet.
\subsubsection{Observational constraints on disk C/O ratios}
\label{coconstraints}
Interesting recent observational results have provided new insights into the elemental ratios of PPDs. Both \citet{vandermarel2021co} both and \citet{bosman2021mapsco} utilised observations of \ce{C2H} and CO gas in PPDs to infer C/O ratios in the disks as a function of radius. Such insights are important. Not only do they constitute a basis of comparison for modelling disk midplane chemistry, they also emphasize the need to have gas and grain dynamical evolution calculated simultaneouly with chemical evolution \citep[as mentioned explicitly in][]{vandermarel2021co}.

Zooming out from a focus on a single PPD in each observational case, it may also be relevant to consider that the global elemental ratios of a PPD should reflect the C/O ratio of its host star. The C/O ratio of the Sun is $\sim$0.55 \citep{asplund2009}, but other stars nearby feature different ratios, possibly ranging from as low as 0.2 up to 0.8 (or close to 1), as shown in \citet{nissen2014,delgadomena2021ctoo}. These different ratios mean that more or less individual carbon, oxygen, nitrogen and sulphur, etc., is available for chemical evolution in a given PPD, and for different disks, different evolved chemical compositions may result. It might therefore be logical if the observed C/O ratios in PPDs be evaluated in light of the host stars' C/O ratios, rather than the solar value. The effects of varying the global C/O ratios and metallicities have been explored in planet formation models by, e.g., \citet{bitschbattistini2020}, although this work used iceline chemistry and thus did not involve chemical reactions.

\section{Next steps and conclusion}

This section will outline some proposed next steps on both the modelling and the observational aspect of this topic, which, in the author's view, will be beneficial to advancing this field in the short term.
\subsection{Theory and modelling}

As discussed in Sect. \ref{connect1} and \ref{connect2}, the efforts to integrate physical and chemical evolution with each other when modelling planet formation has come far. However, they have not merged yet, and there remains some immediate challenges, which need attention if the field is to advance and make the modelling more realistic. These challenges are pointed out below, in no particular order, but starting with the efforts to implement chemical into physical simulations:
\begin{itemize}
    \item Planet formation modelling efforts that utilise iceline chemistry should be conscious and explicit about the drawbacks of this approach. To the extent that it is feasible and doable, researchers should attempt to implement chemical kinetics into the planet formation models. This could possibly be done, either by running chemical and physical evolution simultaneously/in parallel (self-consistently or otherwise) or by post-processing results from chemical evolution models than have already been run.

    The \textsc{KROME} package has made it easier for non-experts in astrochemistry to utilise chemical kinetics, but the astrochemical community should also engage with planet formation modellers, and offer their insights. The astrochemical community will hopefully take the opportunity to collaborate and offer assistance to planet formation modeller in evaluating the degree of astrochemical complexity that is relevant and needed in a given situation, and suggestions on where to implement chemical evolution in a planet formation modelling setup. The goal should be to improve the implementation of astrochemistry into planet formation models, and to guide planet formation modellers in understanding which chemical effects lead to what chemical evolution, and which effects may be more relevant for their specific challenges.

    \item The astrochemical community has developed some important chemical codes that can be used to model chemical evolution in a planet formation context. The community should continue to expand the codes with new chemical species and new reactions and test the outputs of the codes against observational evidence. At the same time, chemical codes should be updated when new constraints on reaction rate coefficients and binding energies from laboratory experiments become available.

    The astrochemical community should explore ways to make chemical kinetics codes faster, and easier to understand and use. Especially the ease-of-use element may be important for how likely planet formation modellers will be to adopt and implement chemical kinetics into their codes. In other words, in order to advance our understanding of the ``chemistry of planet formation'', the astrochemical community should work and collaborate with planet formation modellers.
\end{itemize}

One overarching risk in the fields of astrochemistry and planet formation is that attempts may be made to push a new effect forward (physical or chemical), whilst keeping the other parts of the model very simplified. This happens, e.g., when a planet formation modeller uses sophisticated dust and gas dynamical evolution in the code, but uses iceline chemistry to account for chemical composition, thus assuming no chemical reactions to happen. Likewise, when an astrochemical modeller uses a sophisticated chemical network with various reactions types, including grain-surface ice effects, but uses constant grain sizes, and assumes that grains and gas stay together over planet formation timescales, thus assuming no grain evolution. Each of these cases have strong caveats. Together, the fields would benefit greatly from these two modellers working together, possibly compromising some of each code's physical/chemical complexity, for the benefit of making the merged code run well.

\subsection{Observational}

The \emph{Hubble Space Telescope} (\emph{HST}) and ESO's Very Large Telescope (VLT) have both contributed to constraining exoplanet atmospheric chemistry in the last decade. Alongside, the Atacama Large (sub)Millimeter Array (ALMA) has been, and is still pushing new frontiers in our understanding of the environments in which exoplanets form: protoplanetary disks. While ALMA will likely continue to uncover new insights into PPD chemistry and composition in the near future, both HST and VLT have been pushed to the edge of their observational capabilities with regards to observations of exoplanet atmospheres (the VLTI \textsc{GRAVITY} experiment notwithstanding).

Recently, the ALMA Large Program \emph{Molecules with ALMA at Planet-forming Scales (MAPS)}\citep{oberg2021maps} published results on abundances of more than 20 chemical species towards five PPDs: IM Lup, GM Aur, AS 209, HD 163296, and MWC 480. These results included \citet{legal2021mapsco,bosman2021mapsco}, where the gas-phase C/O ratios in these disks were inferred to be super-stellar. That is, the gas in the planet-forming regions of these disks have a higher carbon-to-oxygen ratios than their respective host stars. These insights are crucial, not only for understanding the C/O ratio of planet-forming gas, but also for guiding and constraining the chemical modelling efforts in connection to planet formation (e.g., a quick comparison to Fig. \ref{eistrupcoevol} shows that, under the modelling assumptions used for that figure, the disk midplane gas C/O ratio remains super-stellar at all radii \emph{only} for evolution times $<$1 Myr).

The most precise observational determination of C/O ratios of exoplanet atmospheres are C/O$_{\mathrm{HR 8799e}}=0.60_{-0.08}^{+0.07}$ for HR 8799e\citep{molliere2020} (similar to C/O ratio of stellar host star HR 8799) and C/O$_\mathrm{\beta Pic b}=0.43\pm 0.05$ for $\beta$ Pic b, see \cite{gravity2020betapic} (the C/O of the stellar host star $\beta$ Pic was yet unknown, when that study was published). Both exoplanets were observed with VLT instruments, including the \textsc{GRAVITY} experiment, and both resulted from exoplanet atmospheric retrieval runs with the \textsc{petitRADTRANS} code by \citet{molliere2019petitradtrans}. Similarly, the \textsc{IGRINS}-instrument at the Gemini-South Observatory was used to determine the \ce{H2O} and CO volume mixing ratios in the day-side hemisphere of the transiting exoplanet WASP-77Ab\citep{line2021ctoo}. These mixing ratios were subsequently used to deduce a C/O ratio of $0.59\pm0.08$ for this day-side hemisphere. With the WASP-77A stellar C/O ratio recently\citep{reggiani2022} inferred to be C/O$_{\mathrm{WASP-77A}}=0.44^{+0.07}_{-0.08}$, this implies that WASP-77Ab has a super-stellar C/O ratio, and \citet{reggiani2022} suggested that this points to this exoplanet having formed exterior to the \ce{H2O} iceline in its natal protoplanetary disk. In the original work by \citet{line2021ctoo}, they inferred that the WASP-77A stellar C/O ratio was similar to that of the WASP-77Ab exoplanet (C/O$_{\mathrm{WASP-77A}}\sim0.55$), and here the conclusion was that the exoplanet had formed inside the \ce{H2O} iceline in its protoplanetary disk.

For the purpose of inferring planetary formation histories from characterised exoplanets, it is therefore of crucial importance to measure elemental ratios like C/O of \emph{both} exoplanet and host star. Stars in the solar neighborhood generally vary in measured C/O ratios from 0.2--0.8 (see \citet{nissen2014ctoo,delgadomena2021ctoo}), and therefore a comparison between a measured exoplanetary C/O ratio and the solar ratio of 0.55 does not carry much meaning, since the Sun might not have the same C/O ratio as the exoplanet's own host star. What holds the potential to distinguish between different formation scenarios for exoplanets is therefore the comparison between exoplanet and host star elemental ratios.

Additionally, this sample of three exoplanets with uncertainties on measured C/O ratios  provides the interesting insight that the atmosphere of $\beta$ Pic b has a distinctly different chemical makeup from HR 8799e and the day-side of WASP-77Ab, which are, in turn, similar to each other. The atmosphere of $\beta$ Pic b must therefore have formed from material that contained relatively more elemental oxygen than carbon, compared to the two other exoplanets. While it is very encouraging that the precision of measurements are now such that the fields of planet formation chemistry and exoplanet atmospheric compositions are now both awaiting the \emph{James Webb Space Telescope (JWST)}.

It is possible that \emph{JWST} will provide a quantum leap in our understanding of exoplanet atmospheric compositions, and planet formation chemistry, alike what ALMA has provided for our understanding of PPDs. The near future with \emph{JWST}, with a possible mission lifetime of 20 years, may thus hold crucial new insights for understanding planet formation chemistry over the next decade. For planet formation chemistry, the \emph{JWST} will enable observations of ices in disks, and also chemical characterisation of the atmospheres of at least $\sim$88 exoplanets\footnote{Number of unique exoplanets targeted for atmospheric spectroscopic characterisation as described in Public PDFs of ERS, GTO, and Cycle 1 GO programs on Space Telescope Science Institute's \emph{JWST} portal (``Exoplanet''-category).}.

On the longer term, the ESA \emph{Ariel} mission is scheduled for launch in 2028. This mission is designed specifically for exoplanet atmospheric characterisation, and may in this way take our understanding even further than \emph{JWST} will. Finally, ground-based facilities are also underway. The ESO Extremely Large Telescope (ELT) in Chile, which will feature a 39 meter diameter mirror, will be humankind's biggest yet optical eye on the sky. ELT will enable, not only very precise chemical characterisation of exoplanet atmospheres, but it may even taken such characterisation to a level, where signatures of life on exoplanets may be detectable (see \citet{schwieterman2018,lopezmorales2019} and references therein).

\subsection{Conclusion}

This review has described and discussed the state-of-the-art for modelling the astrochemistry of planet formation. It was the intention of this review to provide the field of planet formation modelling, the field of astrochemical modelling of PPDs, and the field of exoplanet atmospheric characterisation with an overview of which efforts have been made, and where, in the author's opinion, important new insights have been uncovered, and where work could be done to improve the modelling efforts.

The review has had a strong focus on chemical modelling, and how it relates to disk chemistry and planet formation modelling. The intent of this has, especially, been to provide the planet formation modelling community with insights into what disk chemistry is, which chemical effects may be important and under what physical conditions, and how they can utilise chemical kinetics themselves to make the treatment of chemistry more realistic in their models.

It is the author's opinion that, in order to advance the field of planet formation and the field of astrochemistry, it would be beneficial to work on solutions to merge these two types of modelling into a field of astrochemistry of planet formation. With this, the author wants to strongly encourage, not only collaboration between the fields, but also that researchers in each field offer help, assistance and guidance on best practices and insights into \emph{appropriate} simplifying assumptions, to the other field. This can be achieved, partly through interactions at meetings, partly through reaching out to, e.g., some of the researcher behind work references in this review.

Existing work and future efforts suggest that the astrochemistry of planet formation will soon be better understood.

\paragraph{Acknowledgements.} The author thanks Martin Cordiner, Christopher Bennett, and Eric Herbst from ACS Earth and Space Chemistry under the ACS Biochemistry Division for the invitation to contribute this review. The author also thanks the two anonymous reviewers, whose comments and corrections improved the manuscript.

\bibliographystyle{apalike}
\bibliography{bib2.bib}

\end{document}